\documentstyle[preprint,eqsecnum,aps,epsf]{revtex}

\begin{document}
\preprint{INPP-UVA-01/96}

\title{$F/D$ Ratio in Hyperon Beta Decays\\ 
and\\
Spin Distribution in the Nucleon\\}

\author{X. Song$^*$, P. K. Kabir and J. S. McCarthy\\
{\it Institute of Nuclear and Particle Physics,}\\ 
{\it Department of Physics, University of Virginia,} \\
{\it Charlottesville, VA 22901, USA.}}

\maketitle
\begin{abstract}
It is shown that hyperon beta decay data
can be well accommodated within the framework of 
Cabbibo's SU(3) symmetric description if one allows 
for a small SU(3) symmetry breaking proportional to 
the mass difference between strange and nonstrange 
quarks. The $F/D$ ratio does not depend sensitively
on the exact form of the symmetry-breaking, and the 
best fits are close to the value previously used in 
the analysis of deep inelastic scattering of electrons 
or muons on polarized nucleons. The total quark helicity
and strange quark polarization in the nucleon are discussed.

\end{abstract}
\bigskip
\bigskip

\pacs{13.30.Ce; 14.20.Dh; 11.30.Hv\\
$^*$E-mail address: xs3e@virginia.edu}

\widetext

\section{Introduction}

The spin-dependent (Gamow-Teller) matrix-elements, for 
transitions between members of the baryon octet \cite{gs84}
acquired renewed interest after measurements were made of 
the deep inelastic scattering (DIS) of polarized leptons 
by polarized protons and neutrons 
\cite{emc88,emc93,smc,e142,e143a,e143b}, which provided 
valuable information about the spin structure of the nucleon. 
One of the most important quantities measured in polarized 
DIS is the longitudinal spin structure function $g_1$. In 
the quark parton model, the spin structure function $g_1$ 
is directly related to the quark spin densities: 
$\Delta u(x)$, $\Delta d(x)$, $\Delta s(x)$ etc. where 
$\Delta q(x)\equiv q_{\uparrow}(x)-q_{\downarrow}(x)+
{\bar q}_{\uparrow}(x)-{\bar q}_{\downarrow}(x)$. 
 
To deduce the various quark spin densities from the $g_1$ 
data, one usually assumes that baryons may be assigned to 
a SU(3) flavor octet and uses the relation between the 
quark spin densities and weak matrix elements $F$ and $D$ 
from hyperon semileptonic decays. By using the earlier 
$F/D$ value, the EMC data led to the unexpected conclusion 
\cite{emc88} that the quarks carry at most a small part 
of the spin of either nucleon and furthermore, that there 
is a significant contribution from ``strange'' quarks, 
which necessarily come from the ``sea'' of quark-antiquark 
pairs. This has led to many different suggestions for 
resolution of what has come to be called the ``spin crisis'' 
\cite{aee95,jaffe95,close95}. Among these 
is a suggestion \cite{lipkin89,schafer88} that the 
conclusions may be distorted because the $F/D$ value 
obtained from the hyperon semileptonic decays are based 
on exact SU(3) flavor symmetry. SU(3) symmetry breaking 
effects may significantly change the value. 

There are many works attempting to evaluate the 
SU(3)-breaking effects in the bag model or in the
quark model, by applying center-of-mass corrections 
\cite{bar84,bs86,dhk87}, or by including one-gluon 
exchange interactions \cite{hm88,carlson88} or both 
\cite{ytkk89}. The size of the corrections depends
on the model and assumptions used to describe the
symmetry breaking effects and on the `existing' data
to be fitted. Some authors used their own data and 
concluded \cite{bou83} that there is no signal for the 
breakdown of Cabbibo's SU(3) symmetric description. 
According to Ref.\cite{dhk87}, however, an overall fit
to the existing data using broken SU(3) scheme is
better than that from the assumption of perfect SU(3)
symmetry. Another approach, using the chiral 
effective lagrangian for baryons, \cite{jm91} calculated 
SU(3) symmetry breaking corrections to axial currents 
of the baryon octet arising from meson loops. The size of 
corrections was found to be surprisingly large (the loop 
correction is almost as large as the lowest order result) 
which should already have raised suspicion. In a subsequent
paper \cite{jm91}, including the spin-3/2 baryon decuplet in 
the intermediate state, the meson loop correction to 
the axial currents is significantly reduced but still 
substantial ($\simeq 30-50\%$). However, corrections 
due to higher baryon resonances, which in principle should 
be included in the intermediate states, have been ignored 
in the calculation and may change the result still further.
Thus it appears that the validity of Cabbibo's SU(3) 
symmetric description is far from settled. Most recently, 
instead of model-dependent calculations, an approach 
\cite{es95} based on phenomenological analysis of hyperon 
beta decay data has been suggested to estimate the SU(3) 
symmetry breaking effects. The authors present evidence 
for a strong variation of the $F/D$ parameter between 
various transitions. 

In section II, we consider another approach based on a 
general discussion \cite{gatto64} of SU(3) flavor 
symmetry and its possible breaking and show that the 
hyperon beta decay data are adequately represented by 
at most a small deviation from of Cabibbo's SU(3) 
symmetric description, which can be well accommodated 
within the framework of the usual assumption of a small 
SU(3) breaking proportional to the mass-difference 
between strange and nonstrange quarks. In section III, 
the consequences for the quark spin distribution in 
the nucleon are discussed. A brief summary is given 
in section IV.

\section{SU(3) Symmetry-breaking Effects}

In the quark model, which provides an explicit realization 
of Cabibbo's theory connecting strangeness-conserving and 
strangeness-changing weak interactions, the primary weak 
current responsible for transitions between hadrons is:
$$j^{\mu}_{W}={\bar q}{{\lambda_{W}}\over 2}\gamma^{\mu}
(1+{\gamma_5})q
\eqno (1)$$
with 
$$\lambda_{W}=[\lambda_1 +i\lambda_2]{\rm cos}{\theta_c}
+[\lambda_4 +i\lambda_5]{\rm sin}{\theta_c}
$$
where $\lambda_i$ (i=1,2,...8) denote the Gell-Mann 
matrices and $q$ represents the triplet ( $u$, $d$, $s$ ) 
of basic quark fields. Eq.(1) requires that weak 
transition-elements necessarily transform as a 
component of an SU(3) octet. If baryons are assigned 
to a SU(3) octet, represented in matrix form by :
$$B_a^b=\left( \begin{array}{c}
{1\over {\sqrt 2}}{\Sigma}^o+{1\over {\sqrt 6}}{\Lambda}^o\qquad\quad   
{\Sigma}^+ \qquad\qquad  p \\
{\Sigma}^-\qquad\qquad  -{1\over {\sqrt 2}}{\Sigma}^o+{1\over {\sqrt 
6}}{\Lambda}^o \qquad n  \\ 
{\Xi}^-\qquad\qquad\qquad {\Xi}^o \qquad\qquad -{2\over {\sqrt 6}}{\Lambda}^o
\end{array}\right)
\eqno (2)$$
the SU(3)-octet matrix elements between baryons can be written, 
in the symmetric limit (we are concerned only with the values 
for $q^2\rightarrow 0$, i.e. zero four-momentum transfer ), as
$$DTr({\bar B}\{\lambda_W,B\}_+)+FTr({\bar B}[\lambda_W,B]_-)
\eqno (3)$$
which can also be written as $a_0Tr({\bar B}B\lambda_W)+
b_0Tr({\bar B}\lambda_WB)$, with $a_0=D-F$ and $b_0=D+F$.
However, the SU(3) flavor symmetry is only approximate for 
strangeness-changing processes. If SU(3) 
symmetry-breaking effects cannot be ignored, the 
expressions for the matrix-elements must be generalized. 

We assume that the breaking of SU(3)-flavor symmetry is 
due to a term which transforms like the eighth generator 
of SU(3). This would be the case, for example, if SU(3) 
breaking arose entirely from a mass-difference between 
strange and (degenerate \cite{karl94}) non-strange quarks. 
To first order in the symmetry-breaking interaction, 
transforming like $\lambda_8$, the most general SU(3) 
structure of the weak matrix-elements between baryons 
can be written as 
$$a_0Tr({\bar B}B\lambda_W)+
b_0Tr({\bar B}\lambda_WB)+
aTr({\bar B}B\{\lambda_W,\lambda_8\}_+)+
bTr({\bar B}\{\lambda_W,\lambda_8\}_+B)$$
$$\qquad\qquad +dTr({\bar B}\lambda_8\lambda_WB)+
k[Tr({\bar B}\lambda_W)Tr(B\lambda_8)+
Tr({\bar B}\lambda_8)Tr(B\lambda_W)]/2
\eqno (4)$$
where the first two terms are the ones given in eq.(3) and 
the others are SU(3) symmetry-breaking corrections. The 
corresponding symmetry-breaking parameters $a$, $b$, $d$, 
and $k$ should be small relative to $a_0$ and $b_0$ for 
such a perturbative expansion to be valid. Vector coupling 
constants are not affected to first order \cite{bs60,gatto64}. 
For the ratio of axial-vector to vector amplitudes, eq.(4) 
yields \cite{gatto64} 
$$(G_A/G_V)_{n\rightarrow p}=F+D+2b 
\eqno (5)$$
$$(G_A/G_V)_{\Lambda\rightarrow p}=F+D/3+a/3-2b/3-d/3-k
\eqno (6)$$
$$(G_A/G_V)_{\Sigma^-\rightarrow n}=F-D+a-d
\eqno (7)$$
$$(G_A/G_V)_{\Xi^-\rightarrow \Lambda}=F-D/3+2a/3-b/3+4d/3+k
\eqno (8)$$
where we have listed only those transitions for which 
these ratios are relatively well measured \cite{pdg94}:
$$(G_A/G_V)_{n\rightarrow p}=1.2573\pm 0.0028
\eqno (9)$$
$$(G_A/G_V)_{\Lambda\rightarrow p}=0.718\pm 0.015   
\eqno (10)$$
$$(G_A/G_V)_{\Sigma^-\rightarrow n}=-0.340\pm 0.017 
\eqno (11)$$
$$(G_A/G_V)_{\Xi^-\rightarrow \Lambda}=0.25\pm 0.05 
\eqno (12)$$

Let us first discuss the SU(3) symmetry scheme.
Fig.1 exhibits the results reported in eqs.(9)$-$(12) 
under the assumption that SU(3) symmetry-breaking 
effects are negligible, viz. all breaking parameters 
are zero: $a=b=d=k=0$ in eqs. (5)$-$(8). We see that 
the $(G_A/G_V)$ ratios for the best-measured transitions 
(9)-(11) yield, within the errors, a unique solution 
for $F$ and $D$. While the line corresponding to the 
central value of $(G_A/G_V)$ for the less accurately 
measured $\Xi^-\rightarrow\Lambda$ transition does not 
pass exactly through the same $(F,D)$ point, a downward 
shift of $(G_A/G_V)_{\Xi^-\rightarrow\Lambda}$ by an amount 
equal to the quoted error, is sufficient to bring it into 
agreement with the others. Hence it seems that no 
significant SU(3) symmetry-breaking effect is needed to
describe the existing $(G_A/G_V)$ data. It is also 
interesting to note that the favored 
solution for $F$ and $D$ obtained from data (9)$-$(11) is not 
too different from that predicted by the static SU(6) 
symmetric model with suitable relativistic recoil corrections 
($\simeq$ 25$\%$ reduction\cite{bs86}).

While there does not seem to be any compelling 
evidence demanding the inclusion of SU(3) breaking effects, 
it may be worthwhile to see what is obtained if one takes 
the data, eqs.(9)-(12), at face value and seeks a solution 
allowing any one of the symmetry breaking parameters in 
eq.(4) to be non-zero. We search in the three-dimensional 
space $F$, $D$, $\epsilon$ (where $\epsilon$ denotes one 
of four possible small symmetry breaking parameters: $a$, 
$b$, $d$ or $k$) to find the minimum of the quantity 
$\chi^2$. The results are listed in Table I.

Table I: One-parameter fit
$$
\offinterlineskip \tabskip=0pt 
\vbox{ 
\halign to 1.0\hsize 
   {\strut
   \vrule#                         
   \tabskip=0pt plus 30pt
 & \hfil #  \hfil                  
 & \vrule#                         
 & \hfil #  \hfil                  
 & \vrule#                         
 & \hfil #  \hfil                  
 & \vrule#                         
 & \hfil #  \hfil                  
 & \vrule#                         
 & \hfil #  \hfil                  
 & \vrule#                         
 & \hfil #  \hfil\quad             
   \tabskip=0pt                    %
 & \vrule#                         
   \cr                             
\noalign{\hrule}                   
&       &&b,d,k=0   && a,d,k=0  &&  a,b,d=0  && a,b,k=0 && exp. &\cr 
&       &&a=$-$0.0024 && b=0.0027 &&  k=0.0123 && d=0.0297&& [24]&\cr 
\noalign{\hrule}
&F    && 0.4581  && 0.4576  && 0.4610  && 0.4721&&     &\cr 
&D    && 0.7992  && 0.7943  && 0.7963  && 0.7852&&     &\cr  
&F/D  && 0.573   && 0.576   && 0.579   && 0.601&&      &\cr  
&$(G_A/G_V)_{n\rightarrow p}$
  && 1.2573  && 1.2573  && 1.2573  && 1.2573  && 1.2573$\pm$0.0028&\cr  
&$(G_A/G_V)_{\Lambda\rightarrow p}$
  && 0.723   && 0.721   && 0.714   && 0.724   && 0.718$\pm$0.015 &\cr 
&$(G_A/G_V)_{\Sigma^-\rightarrow n}$
  &&$-$0.343  && $-$0.337  && $-$0.335  && $-$0.343   
&&$-$0.340$\pm$0.017&\cr  
& $(G_A/G_V)_{\Xi^-\rightarrow\Lambda}$
  && 0.190  &&  0.192  &&  0.208  &&  0.250   && 0.25$\pm$0.05 &\cr 
\noalign{\hrule}
&${\chi}^2$  && 1.61   &&  1.42   && 0.86     && 0.20 &&       &\cr 
\noalign{\hrule}
}}$$
\smallskip
\bigskip 

As expected,
it takes only a small non-zero value of any of these, to 
obtain a statistically satisfactory solution. The fifth 
column, with a $d$-type correction, shows the best agreement 
between the calculated and the measured $(G_A/G_V)$ ratios, 
and provides the only indication that inclusion of SU(3) 
breaking effects may be required. The best fits under the 
assumption that SU(3) symmetry breaking arises from terms 
of the type $a$ or $b$, yield values which, in view of 
the quoted errors, are indistinguishable from zero, i.e. 
do not call for any correction at all. Similarly, the 
evidence for non-zero $k$ is marginal. 

From the results listed in Tables I, 
taking average of all the results, we obtain
$$<F>=0.462,\qquad <D>=0.794,\qquad
<F/D>=0.582\qquad 
\eqno (13)$$
These values are consistent with those previously used 
in the analysis of deep inelastic scattering on polarized
nucleons \cite{close93}: 
$$F=0.459\pm 0.008,\qquad D=0.798\pm 0.008\qquad
F/D=0.575\pm 0.016\qquad 
\eqno (14)$$

For illustration, Fig. 2 shows the best fit for 
$k$-type solution. Comparing Fig. 2 and Fig. 1, one sees 
that after inclusion of SU(3) breaking in Cabbibo's scheme,
the lines corresponding to $\Lambda\rightarrow p$ and
$\Sigma^-\rightarrow n$ are both slightly shifted up and 
the only significant change is for the line corresponding
to $\Xi^-\rightarrow\Lambda$. All lines now intersect at 
one point which gives a unique solution of $F$ and $D$ 
for a given parameter set. Similar discussion can be carried 
out for $a$-, $b$- and $d$-type solutions.

It may be noted that all SU(3) symmetry-breaking 
parameters considered in this paper are significantly 
smaller than the SU(3) symmetric parameters $F$ and $D$. 
Compared to the result given in \cite{es95}, our $F/D$ 
value for a given symmetry breaking parameter set is unique 
for the known baryon decay modes. It suggests that the entire 
pattern of existing hyperon semileptonic decay data can 
be very well described in a framework which is basically 
SU(3) flavor symmetry with small SU(3) symmetry-breaking 
effects. Therefore no evidence of $strong$ 
violation for SU(3) symmetry in hyperon beta decay 
data can be found.

\section{Quark Spin Distributions in the Nucleon}

As we mentioned in the introduction, the quark 
spin distributions deduced from the $g_1$ data depend
on the $F/D$ ratio. In the QCD corrected quark parton 
model, we have

$$\Gamma_1^p\equiv 
\int_0^1g_1^p(x)dx={{C_{NS}}\over {18}}
\left[2\Delta u-\Delta d-\Delta s\right] 
+{{C_{S}}\over {9}}\Delta\Sigma
\eqno (15)$$
where $\Delta u=\int_0^1\Delta u(x)dx$ and 
$\Delta\Sigma=\Delta u+\Delta d+\Delta s$ represents the 
fraction of the proton spin carried by all the quarks and 
antiquarks, i.e. the net total quark helicity. 
where $C_{NS}=1-y-3.5833y^2-20.2133y^3-O(130)y^4$ and 
$C_S=1-y/3-0.5495y^2-O(2)y^3$, with $y\equiv\alpha_s/\pi$, 
are QCD correction coefficients for nonsinglet and singlet 
terms \cite{larin}. To simplify the notation, 
we have omitted the variable $Q^2$ in the quantities
listed above. It should be noted that the anomalous 
gluon contributions \cite{gluon} and higher twist effects 
\cite{ht} are not included in (15). The former is 
still a subject of debate and the latter is expected 
to be only a small correction at the low $Q^2$ value
(for example, the E142 $\Gamma_1^n$ data). Combining 
(15) and the following two relations
$$({{G_A}\over {G_V}})_{n\rightarrow p}=F+D=\Delta u-\Delta d
\eqno (16)$$
$$(G_A/G_V)_{\Sigma^-\rightarrow n}=F-D=\Delta d-\Delta s 
\eqno (17)$$
one obtains
$$\Gamma_1^{p(n)}=
{{C_{NS}}\over {12}}(G_A/G_V)_{n\rightarrow p}\left[
+(-)1+{{R-1/3}\over {R+1}}\right] 
+{{C_s}\over 9}\Delta \Sigma 
\eqno (18)$$
hence the data $\Delta\Sigma$ and $\Delta s$ deduced
from $\Gamma_1^p$ depend on $F/D$ value used 
as input in (18). 

Using $(G_A/G_V)_{n\rightarrow p}=1.254\pm 0.006$,
$F/D=0.632\pm 0.062$, and $\alpha_s=0.27$ the EMC data 
\cite{emc88} $(\Gamma_1^p)_{exp}=0.126\pm 0.018$ led to
$$\Delta\Sigma=0.12\pm 0.17,\qquad 
\Delta s=-0.19\pm 0.06
\eqno (19)$$
However, if instead, using $<F/D>=0.582\pm 0.008$ and
the same $C_{NS}=1-\alpha_s/\pi$ and $C_S=1-\alpha_s/3\pi$ 
as used in the EMC analysis \cite{emc88}, one obtains 
$$\Delta\Sigma=0.14\pm 0.17,\qquad 
\Delta s=-0.15\pm 0.06
\eqno (20)$$ 
One can see that by using a smaller $<F/D>$ value, 
$\Delta\Sigma$ increases and the magnitude of 
$\Delta s$ decreases. However, in contrast to the
change of $\Delta s$, the total quark helicity 
$\Delta\Sigma$ is not sensitive to $<F/D>$. This
is consistent with the result given by Lipkin 
\cite{lipkin95}. On the other hand, if we use $C_{NS}$ 
up to $(\alpha_s/\pi)^4$ and $C_S$ up to $(\alpha_s/\pi)^3$ 
as given in \cite{larin}, then (20) becomes 
$$\Delta\Sigma=0.19\pm 0.17,\qquad 
\Delta s=-0.13\pm 0.06
\eqno (21)$$ 
Comparing (21) with (20), one sees that $\Delta\Sigma$
significantly increases after inclusion of higher 
order QCD radiative corrections, which are very important 
in spin analysis, especially at moderate $Q^2$ range 
where the experiments performed. 

Most recently, E143 group obtained more accurate
data of $g_1^p$ which gives 
$\Gamma_1^p=0.125\pm 0.003$ \cite{tj95} with 
$\alpha_s=0.35$. From this, one obtains
$$\Delta\Sigma=0.27\pm 0.04,\qquad \Delta s=-0.10\pm 0.02.
\eqno (22)$$
The difference between the central values of 
$\Delta\Sigma$ (and $\Delta s$) in (22) and in (21) 
is due to that the data are taken at different $Q^2$
and they have different QCD correction coefficients
$C_{NS}(Q^2)$ and $C_{S}(Q^2)$. In obtaining (22), 
$\alpha_s=0.35$ has been used, but for (21) 
$\alpha_s=0.27$ was used. However, considering that 
the errors in (21) are quite large, the results given 
in (22) and (21) are consistent within the errors. 

To avoid possible ambiguity caused by $SU(3)$
symmetry breaking effects, we may choose to only use 
the SU(2) symmetry result (16) and do not use (17).  
From (15) and (16), one can obtain a relation 
between $\Delta\Sigma$ and $\Delta s$
$$c_1\Delta\Sigma - c_2\Delta s=\Gamma_1^p-c_3
\eqno (23)$$
for the proton and similarly
$$c_1\Delta\Sigma - c_2\Delta s=\Gamma_1^n+c_3
\eqno (24)$$
for the neutron, where
$$c_1={{C_{NS}+4C_S}\over {36}},\quad
c_2={{C_S}\over {12}},\quad
c_3=c_2({{G_A}\over {G_V}})_{n\rightarrow p}
\eqno (25)$$
Actually, (23) and (24) are not independent, because
the Bjorken sum rule
$$\Gamma_1^p-\Gamma_1^n=2c_3
={{C_S}\over 6}({{G_A}\over {G_V}})_{n\rightarrow p}
\eqno (26)$$
Therefore one can not deduce the $\Delta\Sigma$ and 
$\Delta s$ separately, even we have both $g_1^p$ 
and $g_1^n$ data. It should be noted that the data
$\Gamma_1^p$ and $\Gamma_1^n$ from the experimental
measurements may not satisfy (26). Hence
the r.h.s. of (23) can be different from that of 
(24). 

To obtain $\Delta\Sigma$ and $\Delta s$ 
separately, we need another relation between these 
two quantities. This can be obtained from (16) and (17)
$$\Delta\Sigma - 3\Delta s=({{G_A}\over {G_V}})_{n\rightarrow p}
+2({{G_A}\over {G_V}})_{\Sigma^-\rightarrow n}
\eqno (27)$$
Using most recent E143 data 
$\Gamma_1^p=0.125\pm 0.003$ 
and 
$\Gamma_1^n=-0.033\pm 0.008$ \cite{tj95},
one obtains from (23) and (24)
$$\Delta\Sigma-0.518\Delta s=0.325\pm 0.023\qquad 
{E143}\ {\rm proton}\ {\rm data}
\eqno (28)$$
and 
$$\Delta\Sigma-0.518\Delta s=0.394\pm 0.063\qquad 
{E143}\ {\rm neutron}\ {\rm data}
\eqno (29)$$
They are shown in Fig.3 (line {\bf 1} for E143 proton 
data and line {\bf 3} for E143 neutron data, where 
Y$\equiv\Delta\Sigma$ and X$\equiv\Delta s$). If we 
assume that there is no strange quark polarization, 
$\Delta s$=0 as predicted by the naive quark model, 
then $\Delta\Sigma=0.33\pm 0.02$ from the proton data 
and $\Delta\Sigma=0.39\pm 0.06$ from the neutron data. 
They are consistent within the errors (see line $\bf 1$
and line $\bf 2$ in Fig. 3).
However, using SU(3) symmetry result eq.(27) and 
combining data (9) and (11), one obtains 
$$\Delta\Sigma-3\Delta s=0.577\pm 0.034
\eqno (30)$$
which is also shown in Fig.3 (line {\bf 2}). One can 
see that the strange quark polarization would be 
negative. From Fig. 3, one obtains that the range of
$\Delta s$ would be
$$\Delta s=-0.12\rightarrow -0.04
\eqno (31)$$
if the SU(3) symmetry is imposed. 

It should be noted that if one can trust
the earlier $\nu-p$ and ${\bar\nu}-p$ elastic scattering 
data, $\Delta s=-0.15\pm 0.09$ \cite{nu} (which gives 
$\Delta\Sigma\simeq 0.19$ for E143 proton data and 
$\Delta\Sigma\simeq 0.32$ for E143 neutron data), then 
the SU(3) symmetry relation (30) is not necessary. 

\section{Summary}

From a general discussion of SU(3) symmetry and its  
breaking, we show that the hyperon beta decay data
can be well accommodated within the framework of 
the usual Cabbibo's SU(3) symmetric description with
a small SU(3) symmetry breaking proportional to the 
mass difference between strange and nonstrange quarks.
The F/D ratio is not far from the value previously 
used in the deep inelastic scattering analysis. Hence 
the result given by using SU(3) symmetry on hyperon 
beta decays will not be significantly disturbed by 
SU(3) symmetry breaking effects. It implies that the 
total quark helicity is still far below naive quark 
model expectation and the strange quark polarization 
seems to be negative provided the anomalous gluon 
contributions and higher twist effects are neglected.

\vspace{0.2 cm}

After completion of this work, we saw the paper by
Ratcliffe \cite{ratcliffe96} which reached similar 
conclusion about SU(3) breaking.

\vspace{0.2 cm}

{\bf Acknowledgements}

The authors thank P. Q. Hung for helpful discussions in 
early stage of this work. We also thank H. J. Weber for
comments and suggestions. This work was supported by 
the U.S. Department of Energy and the Institute of Nuclear 
and Particle Physics, University of Virginia, USA.

\vspace{0.2 cm}

\vfill\eject

\begin{figure}[h]
\epsfxsize=5.0in
\centerline{\epsfbox{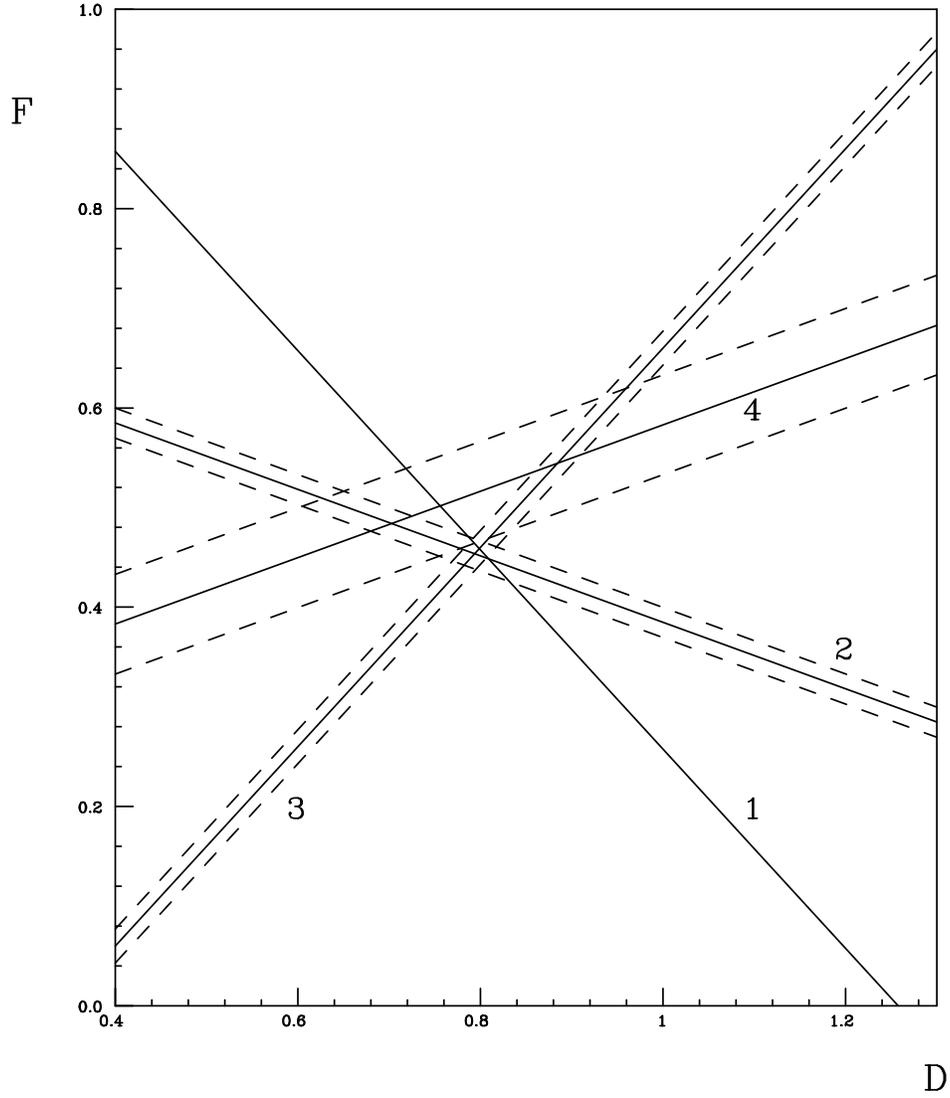}}
\caption{$F-D$ relations determined by experimental values
for various baryonic transitions, assuming no SU(3) breaking. 
Line $\bf 1$: $n\rightarrow p$, line $\bf 2$: $\Lambda\rightarrow p$,
line $\bf 3$: $\Sigma^-\rightarrow n$, line $\bf 4$: 
$\Xi^-\rightarrow\Lambda$.}
\end{figure}

\begin{figure}[h]
\epsfxsize=5.0in
\centerline{\epsfbox{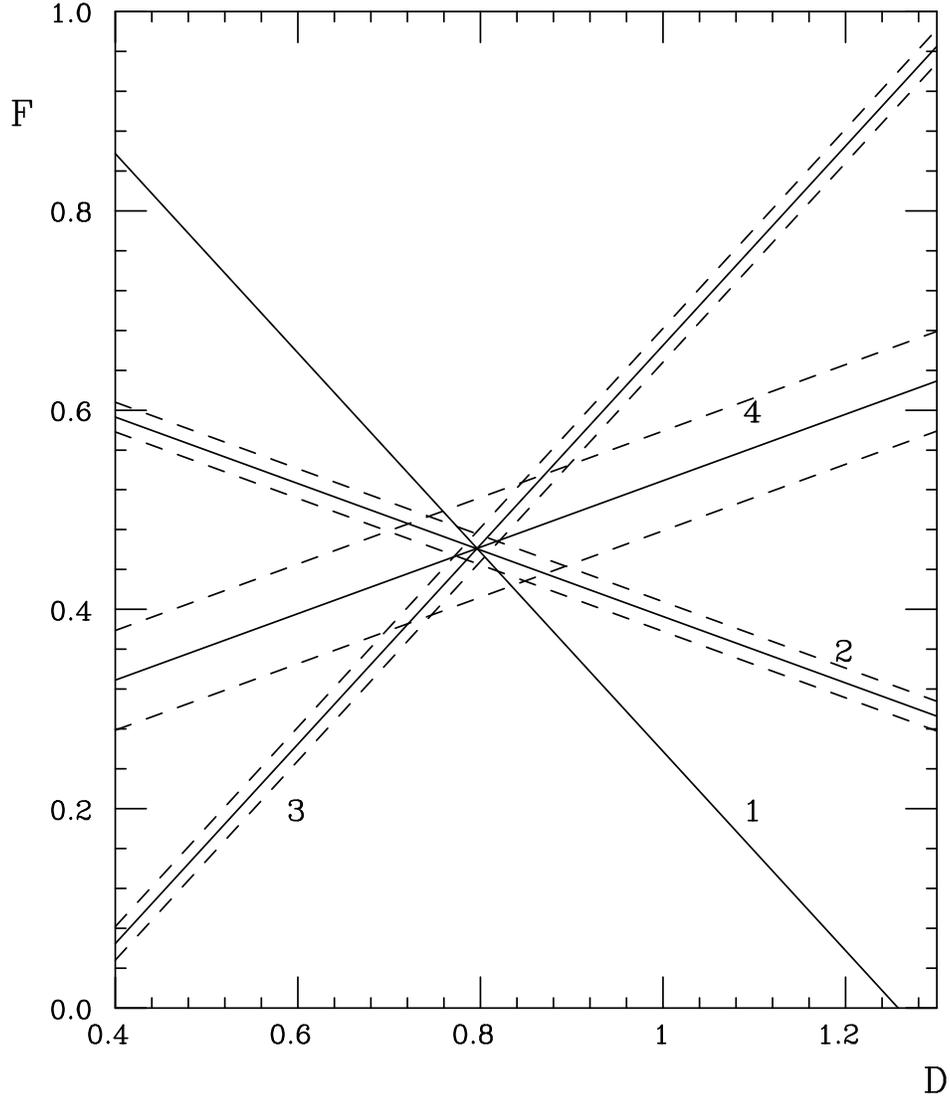}}
\caption{$F-D$ relations, as in Fig.1, allowing for $k$-type SU(3) 
breaking with $k=0.0123$. Line $\bf 1$: $F+D=1.2573\pm 0.0028$, line $\bf 
2$: $F+D/3+k=0.714\pm 0.015$, line $\bf 3$: $F-D=-0.335\pm 0.017$, 
line $\bf 4$: $F-D/3-k=0.208\pm 0.050$.}
\end{figure}

\begin{figure}[h]
\epsfxsize=5.0in
\centerline{\epsfbox{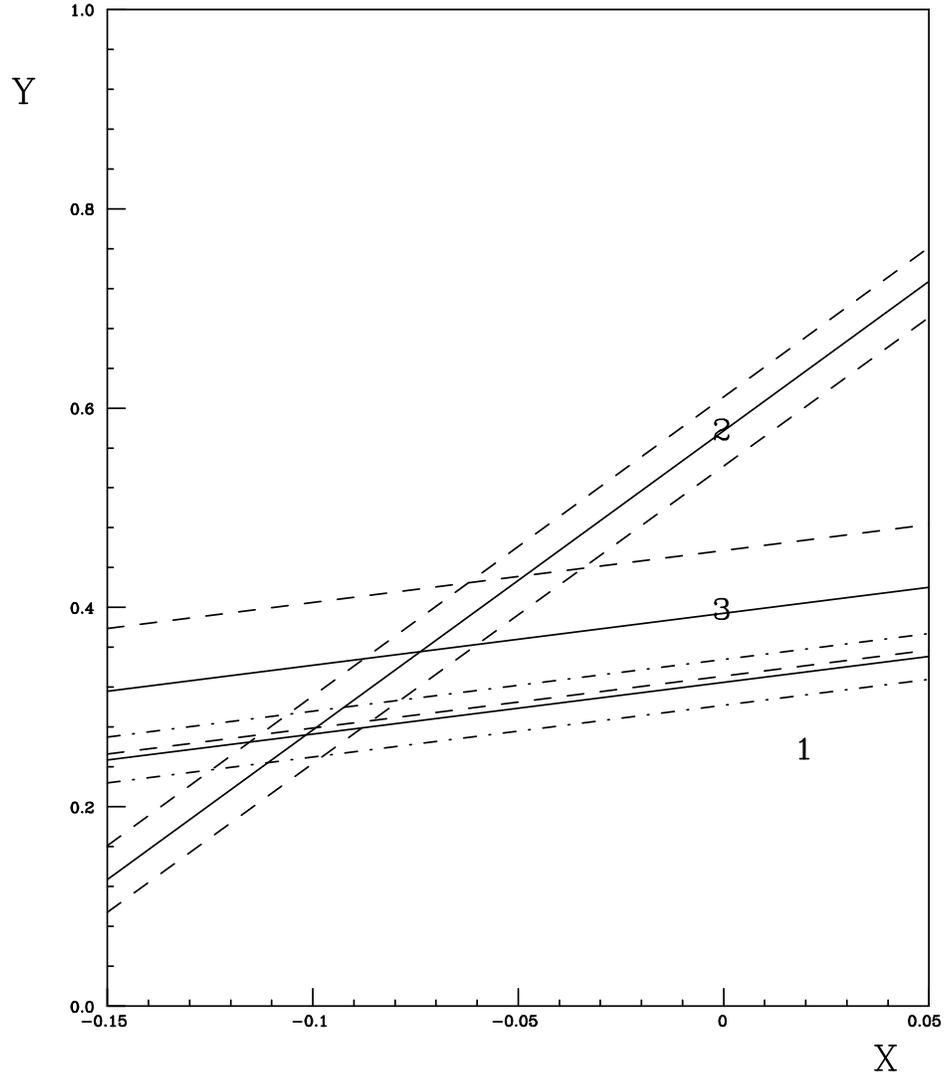}}
\caption{Plot of total quark helicity $Y\equiv\Delta\Sigma$ and
strange quark polarization $X\equiv\Delta s$ constrained by
the E143 proton data (line $\bf 1$: eq.(29)) and neutron data
(line $\bf 3$: eq.(30)), and SU(3) symmetry relation (line
$\bf 2$: eq.(34)).}
\end{figure}



\baselineskip 16pt
\begin{thebibliography}{99}


\bibitem{gs84}
     For an earlier review, see J.-M.~Gaillard and G.~Sauvage, 
{\sl Ann. Rev. Nucl. Part. Sci.} {\bf 34}, 351 (1984).
  
\bibitem{emc88} 
     J.~Ashman et al., {\sl Phys. Lett.} {\bf B206}, 364 (1988); 
Nucl. Phys. {\bf B328}, 1 (1989) 
  
\bibitem{emc93} 
     EMC Collaboration, B. Adeva et al., {\sl Phys. Lett.} {\bf B302},
534 (1993)

\bibitem{smc}
     B.~Adeva {\it et al.} {\sl Phys.Lett.} {\bf B302}, 553 (1993);\ 
     {\bf B320}, 400 (1994)\\
     D.~Adams {\it et al.}, {\sl Phys.Lett.} {\bf B329}, 399 (1994);\
     {\bf B336}, 125 (1994).

\bibitem{e142}
     P.~L.~Anthony {\it et al.} {\sl Phys.Rev.Lett.} {\bf 71}, 959 (1993).

\bibitem{e143a}
     K.~Abe {\it et al.} {\sl Phys.Rev.Lett.} {\bf 74}, 346 (1995);\
     {\bf 75}, 25 (1995).

\bibitem{e143b}
     K.~Abe {\it et al.} {\sl SLAC-PUB-6982,} Sep. 1, 1995.

\bibitem{aee95}
     M.~Anselmino, A.~Efremov and E.~Leader, {\sl Phys. Rep.} 
{\bf 261}, 1 (1995)
    
\bibitem{jaffe95}
     R.~L.~Jaffe, {\sl Physics Today}, September, 24 (1995).

\bibitem{close95}
     F.~E.~Close, hep-ph/9509251, September 8 (1995)

\bibitem{lipkin89}
     H.~Lipkin, {\sl Phys.Lett.} {\bf B230}, 135 (1989).

\bibitem{schafer88}
     A.~Schafer, {\sl Phys.Lett.} {\bf B208}, 175 (1988).

\bibitem{bar84}
     J.~Bartelski et al., {\sl Phys. Rev.} {\bf D29}, 1035 
(1984).

\bibitem{bs86}
     M.~Beyer and S. K. Singh, {\sl Z. Phys.} {\bf C31}, 421 (1986).

\bibitem{dhk87}
    J.~F.~Donoghue, B.~R.~Holstein and S.~W.~Klimt {\sl Phys. Rev.} 
{\bf D35}, 934 (1987).

\bibitem{hm88}
     H.~H$\phi$gaasen and F.~Myhrer, {\sl Phys. Rev.} {\bf D37}, 
1950 (1988).

\bibitem{carlson88}
     L.~J.~Carlson, R.~J.~Oakes and C.~R.~Willcox,
{\sl Phys. Rev.} {\bf D37}, 3197 (1988).

\bibitem{ytkk89}
     T.~Yamaguchi, K.~Tsushima, Y.~Kohyama, and K.~Kubodera, 
{\sl Nucl. Phys.} {\bf A500}, 429 (1989).

\bibitem{bou83}
     M.~Bouquin et al., {\sl Z. Phys.} 
{\bf C12}, 307 (1982);\  {\bf C21}, 1;\ 17;\ 21 (1983).

\bibitem{jm91}
     E.~Jenkins and A.~V.~Manohar, {\sl Phys.Lett.} 
{\bf B255}, 558 (1991);\ {\bf B259}, 353 (1991).

\bibitem{es95}
     E.~Ehrnsberger and A.~Schafer, {\sl Phys.Lett.} 
{\bf B348}, 619 (1995).

\bibitem{karl94}
     For discussion of isospin breaking effects from 
the mass difference between up and down quarks and 
electromagnetic interactions, see G.~Karl, {\sl Phys.Lett.} 
{\bf B328}, 149 (1994).

\bibitem{bs60}
     R.~E.~Behrend and A.~Sirlin, {\sl Phys. Rev. Lett.} 
{\bf 4}, 186 (1960).

\bibitem{gatto64}
     M.~Ademollo and R.~Gatto, {\sl Phys. Rev. Lett.} 
{\bf 13}, 264 (1964).

\bibitem{pdg94}
     Particle Data Group, {\sl Phys. Rev.} {\bf D50}, 1173 (1994)


\bibitem{close93}
     F.~E.~Close and R.~G.~Roberts, {\sl Phys. Lett.} {\bf B316},
165 (1993) 

\bibitem{larin}
     S.~A.~Larin, {\sl Phys. Lett.} {\bf B334}, 192 (1994)

\bibitem{gluon}
     G.~Altarelli and G.~Ross, {\sl Phys. Lett.} {\bf B212}, 391 (1988)\\
     A.~V.~Efremov and O.~V.~Teryaev, {\sl Dubna Report}, JINR-E2-88-287, 
(1988) \\
     R.~D.~Carlitz, J.~D.~Collins and A.~H.~Mueller, {sl Phys. Lett.}
{\bf B214}, 229 (1988)

\bibitem{ht}
     G.~G.~Ross and R.~G.~Roberts, {\sl Phys. Lett.} {\bf B322},
425 (1994)\\
     E.~Stein, P.~Gornicki, L.~Mankiewicz and A.~Schafer,
{\sl Phys. Lett.} {\bf B343}, 369 (1995); {\bf B353}, 107 (1995)

\bibitem{lipkin95}
     H.~Lipkin, {\sl Phys.Lett.} {\bf B353}, 119 (1995).

\bibitem{tj95}
    T.~J.~Liu, PhD. Thesis, University of Virginia, September, 1995

\bibitem{nu}
    L.~H.~Ahrens~et al., {\sl Phys. Rev.} {\bf D35}, 785 (1987).

\bibitem{ratcliffe96}
     P.~G.~Ratcliffe, {\sl Phys.Lett.} {\bf B365}, 383 (1996).


\end{thebibliography}
\end{document}